\documentclass[
    reprint,
    superscriptaddress,
    preprintnumbers,
    nofootinbib,
    nobibnotes,
    aps,
    prd
    ]{revtex4-2}
\pdfoutput=1

\usepackage{graphicx}
\usepackage[bookmarksopen, bookmarksnumbered, bookmarksopenlevel=2, hidelinks]{hyperref}

\hypersetup{
	pdftitle={Open Strings and Heterotic Instantons},
	pdfauthor={Rafael \'Alvarez-Garc\'ia, Christian Knei{\ss}l, Jacob M. Leedom and Nicole Righi}
}

\usepackage{amsmath}
\usepackage{amssymb}
\usepackage{slashed}

\usepackage{orcidlink}
\usepackage{soul}
\usepackage[all]{nowidow}
\usepackage{microtype}
\usepackage[capitalise]{cleveref}

\newcommand{\fstop}{\,.}
\newcommand{\mc}{\mathcal}

\newcommand{\ints}{\mathbb{Z}} 

\newcommand{\HO}{\mathrm{SemiSpin}(32)}

\DeclareMathOperator{\tr}{tr}

\begin{document}

\preprint{DESY 24-114}
\preprint{KCL-PH-TH/2024-44}
\preprint{MPP-2024-157}
\preprint{ZMP-HH/24-15}

\title{Open Strings and Heterotic Instantons}

\author{Rafael \'Alvarez-Garc\'ia\,\orcidlink{0000-0002-8927-2656}}
\email{rafael.alvarez.garcia@desy.de}
\affiliation{II.\ Institut f\"ur Theoretische Physik, Universit\"at Hamburg, Luruper Chaussee 149, 22607 Hamburg, Germany}

\author{Christian Knei{\ss}l}
\email{ckneissl@mpp.mpg.de}
\affiliation{Max-Planck-Institut f\"{u}r Physik (Werner-Heisenberg-Institut), Boltzmannstr.\ 8, 85748 Garching, Germany}

\author{Jacob M.\ Leedom\,\orcidlink{0000-0003-4911-2188}}
\email{jacob.michael.leedom@desy.de}
\affiliation{Deutsches Elektronen-Synchrotron DESY, Notkestr.\ 85, 22607 Hamburg, Germany}

\author{Nicole Righi\,\orcidlink{0000-0002-3127-1773}}
\email{nicole.righi@kcl.ac.uk}
\affiliation{Theoretical Particle Physics and Cosmology Group, Physics Department, King's College London, Strand, London WC2R 2LS, United Kingdom}


\begin{abstract}
    Motivated by closed string perturbation theory arguments by S.\,Shenker, we consider non-perturbative effects of characteristic strength $\mathcal{O}(e^{-1/g_{s}})$, with $g_{s}$ the closed string coupling constant, in supersymmetric critical heterotic string theories. We argue that in 10D such effects arise from heterotic ``D-instantons," i.e.\ heterotic disk diagrams, whose existence relies on a non-trivial interplay between worldsheet and spacetime degrees of freedom. In compactifications of the $\mathrm{SemiSpin}(32)$ heterotic string, we argue that similar effects can arise from wrapped Euclidean non-BPS ``D-strings." Two general principles arise: The first is that the consistency of those heterotic branes on which the fundamental string can end relies on an inflow mechanism for spacetime degrees of freedom. The second is that Shenker's argument, taken to its logical conclusion, implies that all closed string theories must exhibit open strings as well.
\end{abstract}

\maketitle


\section{Introduction}
\label{sec:intro}

Extended objects are integral to our modern understanding of quantum gravity. Fundamental strings define the perturbative corners of string theory, which are further populated by objects of varied spatial extent intimately related to non-perturbative aspects, such as \mbox{D$p$-branes}. Moreover, extended objects play crucial roles in the tightly interconnected web of string dualities and in M-theory.

The $\mathrm{SemiSpin}(32)$ and $\left( \mathrm{E}_{8} \times \mathrm{E}_{8} \right) \rtimes \mathbb{Z}_{2}$ heterotic string theories\footnote{Following~\cite{McInnes:1999va,McInnes:1999pt}, we use $\mathrm{SemiSpin}(32)$ to denote the $\mathrm{Spin}(32)/\ints_2$ quotient that is not isomorphic to $\mathrm{SO}(32)$.} \cite{Gross:1985fr,Gross:1985rr}\,---\,which we denote in what follows as the HO and HE theory, respectively\,---\,stand out among the five 10D superstrings in this regard: They feature the \mbox{F-string}, the \mbox{NS5-brane} and some non-supersymmetric \mbox{$p$-branes} \cite{Polchinski:2005bg,Bergshoeff:2006bs}, of which new types were recently found in \cite{Kaidi:2023tqo}. Their left-right asymmetric worldsheet theory prevents, however, the definition of boundary states within the conformal field theory, meaning that the rich spectrum of \mbox{D$p$-branes} present in Type~II and Type~I string theory is absent.

The significance of D$p$-branes can be underscored by examining string perturbation theory. A universal property of \mbox{D$p$-branes} is that their tensions scale as $\tau_{p} \propto g_{s}^{-1}$, stemming from the fact that the DBI~action describing them is an open string tree-level action computed by disk amplitudes. In a classic paper \cite{Shenker:1990uf}, Shenker argued that, to make sense of the asymptotic series for amplitudes produced in closed string perturbation theory, there should be universal leading non-perturbative corrections of order $\mathcal{O}\left( e^{-1/g_{s}} \right)$. This inspired Polchinski to study D-instantons and disk amplitudes as a source of Shenker effects in the 26D bosonic string~\cite{Polchinski:1994fq}; his arguments can be extended to the Type~II and Type~I theories in 10D. Upon compactification, wrapped Euclidean \mbox{D-branes} give rise to additional Shenker effects.

This amplitudes argument applies to all closed string theories, and hence in particular to the heterotic ones. However, since these lack \mbox{D-branes}, the possible origin of their Shenker effects remains unclear. Early on, Silverstein argued for heterotic $\mathcal{O} \left( e^{-1/g_{s}} \right)$ effects as the S-duals of Type~I worldsheet instantons \cite{Silverstein:1996xp}, but a fundamental understanding of the effects on the HO side is still absent. Moreover, Shenker effects should be present already in the 10D heterotic theories, in which they cannot be S-dual to Type~I worldsheet instantons.

One may be tempted to dismiss Shenker's argument as perhaps only relevant for a subset of theories, but the existence of the effects he predicted in the 10D HO~theory has been demonstrated by Green and Rudra \cite{Green:2016tfs}. Via compactifications of 11D supergravity, they computed the non-perturbative corrections to the HO $R^{4}$-term, concluding that its dilaton-dependent coefficient is given by a real-analytic Eisenstein series, whose Fourier-Bessel expansion is
\begin{equation}
\begin{split}
    E_{\frac{3}{2}} \left( ig_s^{-1} \right) &= 2\zeta(3)g_{s}^{-\frac{3}{2}} + 2\zeta(2)\sqrt{g_s} + \sum_{n\in\ints^+}8\pi\sigma_{-1}(\lvert n \rvert)\\
    &\quad\times \exp(-2\pi\lvert n\rvert/g_s)(1+\mathcal{O}(g_s))\,.
\end{split}
\label{eq:eisenstein}
\end{equation}
Interestingly, they also found the Shenker effects in 10D Type~I, as required by S-duality, but concluded that they are absent from the 10D HE $R^{4}$-term; only after compactifying on $S^{1}$ with an appropriate Wilson line do they find the relevant terms in the HE theory. While the homotopy groups of the heterotic gauge groups hint at the existence of some objects that could explain the results of Green and Rudra, these remain mysterious, and it is not clear how the theory realizes them. Nevertheless, the Shenker effects in the $R^{4}$-term were recently found to be essential for the heterotic theories to align with our expectations on the behavior of the quantum gravity cut-off~\cite{vandeHeisteeg:2023dlw}.

In this \textit{Letter}, we seek to shed some light on the nature of heterotic Shenker effects by describing the existence of heterotic ``D-instantons," i.e.\ heterotic open disk diagrams, by extending Polchinski's idea for open heterotic strings \cite{Polchinski:2005bg}. These are possible thanks to an inflow mechanism arising from fermion zero modes in spacetime. In \cref{sec:inheritance}, we will argue that such objects are inherited from Type~IIB by regarding the HO~theory as its non-perturbative quotient, following \cite{Hull:1997kt,Hull:1998he}. We continue in \cref{sec:K-theory} by expounding their topological properties and connecting them to the relevant cobordism group, as well as concepts in the Swampland Program. \cref{sec:hetdisks} starts by reviewing Shenker's argument and how the predicted effects are realized in many string theories by virtue of the disk diagrams enabled by the open string sector; we argue that such diagrams are also possible in the heterotic theories thanks to the aforementioned inflow mechanism for spacetime fermion zero modes. While the preceding sections are mostly concerned with the HO~theory, we apply these ideas to the HE~theory in \cref{sec:other-heterotic-theories}, and briefly comment on other heterotic string theories. Finally, we summarize and conclude in \cref{sec:conclusions}.

\section{Branes in Type IIB quotients}
\label{sec:inheritance}

The HO theory can be regarded as a non-perturbative orientifold of Type~IIB string theory \cite{Hull:1997kt,Hull:1998he}. Extended objects of a quotient theory can be understood by studying how the ones of its cover theory are affected by the relevant group action; in this section, we analyze the HO~theory from this perspective. Let us commence by recalling some well-known facts about the 10D quotients of Type~IIB string theory.

The non-perturbative duality group of Type~IIB is $G_{\text{non-pert}} = \mathrm{Pin}^{+}(\mathrm{GL}(2,\mathbb{Z}))$ \cite{Hull:1994ys,Pantev:2016nze,Tachikawa:2018njr}. It includes the commonly considered $\mathrm{SL}(2,\mathbb{Z})$-duality acting on the bosonic content of the theory and a subgroup $D_{4} = C_{4} \rtimes \mathbb{Z}_{2}$ of perturbative symmetries, with $\Omega (-1)^{F_{L}}$ the generator of $C_{4}$ and $(-1)^{F_{L}}$ the one of $\mathbb{Z}_{2}$ \cite{Dabholkar:1997zd}. Here $\Omega$ and $(-1)^{F_{L}}$ are the worldsheet parity and spacetime left-moving fermion number operators, respectively. Of particular importance for us will be the S-duality action, corresponding to the $S$ generator of $\mathrm{SL}(2,\mathbb{Z})$, using standard notation, which acts on the axio-dilaton $\tau := C_{0} + ie^{-\Phi}$ as $S: \tau \mapsto -1/\tau$.

The BPS spectrum of Type~IIB contains the \mbox{D-branes} of odd space dimensionality, the \mbox{F-string} and the \mbox{NS5-brane}. One can conclude that S-duality also implies the existence of a second spacetime-filling brane, namely the \mbox{NS9}-brane coupling to the non-dynamical 10-form $B_{10}$ \cite{Hull:1997kt,Bergshoeff:1998re}.

We can now consider quotients of Type~IIB by elements of $G_{\text{non-pert}}$. Those yielding a theory with a supergravity limit and preserving some supersymmetry must, by string universality in 10D \cite{Adams:2010zy}, lead to one of the five superstring theories. There are two such perturbative quotients: the one by $(-1)^{F_{L}}$ and the one by $\Omega$.

Taking the quotient of Type~IIB string theory by $(-1)^{F_{L}}$ we obtain Type~IIA string theory \cite{Sen:1996na,Sen:1996yy,Dabholkar:1997zd}. From the perspective of the Green-Schwarz superstring, $(-1)^{F_{L}}$ acts only on the Grassmannian variables of the target space, meaning that this atypical construction of Type~IIA can be regarded as a superspace orbifold of Type~IIB.

Considering instead the quotient of Type~IIB by $\Omega$ leads to Type~I string theory. The F-string becomes unorientable, signalling the presence of a spacetime-filling \mbox{O9-plane}; an accompanying stack of 32 \mbox{D9-branes} ensures tadpole cancellation in the consistent background. The $\Omega$-even Type~IIB D$p$-branes, i.e.\ those with $p \in \{1,5,9\}$, comprise the BPS spectrum of Type~I. The D$p$-branes with $p \in \{-1,0\}$ are non-BPS stable configurations, as can be seen from a K-theoretic \cite{Witten:1998cd} or a BCFT \cite{Sen:1998ki,Frau:1999qs,Frau:2000hq} analysis. The Type~I D-instanton is of particular interest, corresponding to a D$(-1)$-$\overline{\mathrm{D}(-1)}$ superposition in Type~IIB, for which not only the $(-1)$-$(-1)$ tachyons are projected out under $\Omega$, but also those in the $(-1)\text{-}9 \oplus 9\text{-}(-1)$ sector, see \cite{Witten:1998cd,Frau:1999qs,Frau:2000hq}. The stability of the Type~I D-instanton is associated with the K-theory charge $\mathrm{KO}(S^{10}) = \mathbb{Z}_{2}$, meaning that even numbers of them can annihilate \cite{Witten:1998cd}.

Having exhausted the perturbative quotient constructions descending from Type~IIB, we turn our attention to the non-perturbative part of its duality group. In particular, following the works by Hull \cite{Hull:1997kt,Hull:1998he}, we consider taking the quotient by the operator $\widetilde{\Omega} := S \Omega S^{-1}$. Since $\widetilde{\Omega}$ is obtained by the conjugation action of S-duality on $\Omega$, quotienting the perturbative limit of Type~IIB by it leads, in view of heterotic/Type~I duality \cite{Witten:1995ex,Polchinski:1995df}, to the HO~theory. The Type~IIB D-string becomes unorientable, indicating the presence of the S-dual pair of the \mbox{O9-plane} alongside 32 \mbox{NS9-branes} cancelling the tadpole. While the parallels with the Type~I construction are clear, the role played by S-duality in this quotient construction means that we must abandon the perturbative worldsheet paradigm and regard it as an orientifold of the complete, spacetime theory.

However, this picture can be connected to the perturbative HO~frame as follows. Recall that in the Type~I frame, the \mbox{F-string} can extend between \mbox{D-branes}, in particular between a \mbox{D-string} and the background stack of \mbox{D9-branes}. At finite values of the string coupling $g_{s}^{\mathrm{I}}$, said non-BPS string becomes unstable, with a lifetime inversely proportional to $\sqrt{g_{s}^{\mathrm{I}}}$. Considering the HO~theory at perturbative, but finite values of the coupling $g_{s}^{\mathrm{HO}}$, an analogous picture arises through Hull's orientifold \cite{Hull:1997kt,Hull:1998he}: An unstable non-BPS HO~``D-string" can extend between the F-string and the background \mbox{NS9-branes}, the latter providing it with Chan-Paton factors. This \mbox{D-string} tethers the gauge charges to the fundamental string and, in the strict perturbative limit, completely retracts onto it; its massless spectrum provides 32 left-handed Majorana-Weyl worldsheet fermions transforming under the gauge group, i.e.\ the asymmetry in degrees of freedom of the heterotic worldsheet construction \cite{Hull:1998he}. From the perspective of Hull's orientifold, this inflow of degrees of freedom of the D-string is what distinguishes the heterotic F-string.

In those compactifications of the HO~theory for which the internal space has non-trivial 2-cycles, wrapped \mbox{Euclidean} HO D-strings will lead to instanton corrections that can be identified with Shenker effects. The argument for HO Shenker effects as the S-duals of Type~I worldsheet instantons \cite{Silverstein:1996xp} finds a natural explanation within Hull's orientifold picture, resolving one of the puzzles raised in \cref{sec:intro}. Note that the instability of these configurations does not prevent them from contributing to the path integral; indeed, all saddle points must be summed over in the quantum theory, and D-instantons with tachyonic modes can provide sensible contributions~\cite{Balthazar:2019rnh,Balthazar:2019ypi,Sen:2020cef}.

Earlier, we reviewed how a tower of D-instantons descends from Type~IIB to Type~I, as can be understood from the orientifold construction of the latter. Similarly, Hull's orientifold picture leads us to believe that the HO~theory inherits a tower of ``D-instantons" from Type~IIB as well. These were not discussed in \cite{Hull:1997kt,Hull:1998he}, and are, in fact, more subtle to track in heterotic/Type~I duality than the strings we just examined. Since heterotic/Type~I duality stems from Type~IIB \mbox{S-duality}, we can appreciate why in the cover theory. To regard 10D Type~IIB as an appropriate limit of M-theory on $T^{2}$ we need to make a choice of F- and \mbox{D-string}, i.e.\ a marking of the \mbox{1-cycles} of $T^{2}$. This determines a concrete Type~IIA dual frame and allows us to understand the tower of D-instantons as wrapped D0-particles. \mbox{S-duality} then corresponds to a different marking of the \mbox{1-cycles}, and hence a different pair of Type~IIA/IIB frames with its own tower of wrapped D0-particles/D-instantons. The resummed tower of D-instantons of one Type~IIB frame reorganizes collectively into the one of its S-dual, but the instantons cannot be individually tracked along the process. The appearance of the Eisenstein series in the \mbox{$R^{4}$-term} of Type~IIB showcases how this works: The resummed tower of instantons leads to the Eisenstein series, but individual instantons are only identified after we perform its Fourier-Bessel expansion, i.e.\ after we make a concrete choice of frame.

In spite of this, Hull's orientifold allows us to gain a heuristic intuition about the HO instantons. On the Type~I side, we have a tower of $\mathbb{Z}_{2}$-charged D-instantons probed by the F-string. Slightly deviating from the strict perturbative limit, the F-string starts to retract into the instantons. At strong coupling, we can take the perturbative HO point of view. While the instantons cannot be individually tracked, the instanton tower of the HO~frame can be thought of as a collective reorganization of the Type~I instantons, onto which the HO ``D-string" has completely retracted. Said D-string is also the object tethering the instantons to the background \mbox{NS9-branes}, meaning that we should have $\mathbb{Z}_{2}$-charged HO~instantons with a gauge profile. While heuristic in nature, this argument points towards the properties of the HO~instantons that we will independently motivate in the upcoming sections. These are the sources of Shenker effects that we propose are at play in the 10D HO~theory and that, in particular, explain the results of \cite{Green:2016tfs} for the $R^{4}$-term.

\section{K-theory, cobordism and the Swampland}
\label{sec:K-theory}

The K-theoretic discussion of \cite{Witten:1998cd} argued for the presence of a stable non-BPS Type~I \mbox{$(-1)$-brane} associated with a non-trivial $\mathrm{SO}(32)$-bundle over the compactified \mbox{Euclidean} spacetime $S^{10}$, whose existence is signalled by the non-trivial class in
\begin{equation}
    \pi_{10}(B\mathrm{SO}(32)) = \pi_{9}(\mathrm{SO}(32)) = \mathbb{Z}_{2}\,,
\end{equation}
where $BG$ denotes the classifying space of the topological group $G$. In the HO~theory we have a perturbative $\mathrm{SemiSpin}(32)$-bundle and the non-trivial homotopy group
\begin{equation}
    \pi_{10}(B\mathrm{SemiSpin}(32)) = \pi_{9}(\mathrm{SemiSpin}(32)) = \mathbb{Z}_{2}\,.
\end{equation}
This gauge bundle is supported on \mbox{NS9-branes}, and hence lacks the conventional K-theory interpretation of the Type~II and Type~I theories; we nonetheless examine its K-theoretic properties. It is tempting to associate a purely gauge instanton with the non-trivial class above, but this would be incorrect. An extension of Derrick's theorem~\cite{Derrick:1964ww} implies that a purely gauge configuration for the instanton, which would be characterized by an action scaling like $g_{\mathrm{YM}}^{-2} \sim \left( g_{s}^{\mathrm{HO}} \right)^{-2}$, is untenable: the HO~instanton must be stringy in nature.

We will return to the delicate interplay between the non-trivial gauge configuration and the heterotic strings below, but for the moment let us further expound the topological properties of the HO~instanton. Such properties can be gleaned from the cobordism group
\begin{equation}
    \Omega_{10}^{\mathrm{Spin}}(B\mathrm{SemiSpin}(32)) \cong 10\mathbb{Z}_{2}\,,
\label{eq:SS32-cobordism-group}
\end{equation}
which was calculated in \cite{SS32Cobordism}. Here $\Omega_{n}^{\xi}(X)$ is the group of equivalence classes $[(M,f)]$ of manifolds $M$ endowed with $\xi$-structure and paired with maps $f: M \rightarrow X$ to the topological space $X$. The focus on~\eqref{eq:SS32-cobordism-group} corresponds to our interest in background gauge configurations over the 10D spin manifold representing Euclidean spacetime. We can extract various facts from the non-vanishing of $\Omega_{10}^{\mathrm{Spin}}(B\mathrm{SemiSpin}(32))$.

First, a 9D theory of fermions charged under the group \mbox{$G = \mathrm{SemiSpin}(32)$} will have a global anomaly characterized by the $\eta$-invariant~\cite{Witten:2019bou}. When the theory is defined over a sphere, the anomaly is also characterized by $\pi_{9}(\mathrm{SemiSpin}(32)) = \mathbb{Z}_{2}$ and measured by the mod 2 Atiyah-Singer index theorem~\cite{PMIHES_1969,Atiyah:1970ws,Atiyah:1971rm}. This is a variation of the quintessential example of such a global anomaly, namely Witten's $\mathrm{SU}(2)$ anomaly~\cite{Witten:1982fp}. For us, it is important because such an anomaly heralds, as explained by Witten for the $\mathrm{SU}(2)$ case, the existence of fermion zero modes in one dimension higher, i.e.\ for the gauge instanton background in the HO~theory. These will play a crucial role in \cref{sec:hetdisks}.

Secondly, the K-theory charge of the instanton can be inferred from~\eqref{eq:SS32-cobordism-group}. The Anderson-Brown-Peterson theorem~\cite{Anderson:1967} relates $\Omega_{10}^{\mathrm{Spin}}(B\mathrm{SemiSpin}(32))$ to a direct sum of K-theory groups, among which $\mathrm{ko}_{10}(B\mathrm{SemiSpin}(32))$ is found. This group is, due to a short exact sequence and the splitting lemma, isomorphic to the direct sum $\mathrm{ko}_{10}(\mathrm{pt}) \oplus \widetilde{\mathrm{ko}}_{10}(B\mathrm{SemiSpin}(32))$. We identify that the HO instanton has indeed non-trivial \mbox{$\mathbb{Z}_2$-valued} \mbox{K-theory} charge: A purely gravitational piece corresponding to $\mathrm{ko}_{10}(\mathrm{pt}) = \widetilde{\mathrm{ko}}(S^{10}) = \mathbb{Z}_{2}$, that we study in \cref{sec:other-heterotic-theories}, and a gauge piece associated with \mbox{$\mathbb{Z}_{2} \subseteq \widetilde{\mathrm{ko}}_{10}(B\mathrm{SemiSpin}(32))$}, to which we turn our attention in \cref{sec:hetdisks}. Both groups actually signal a non-trivial mod 2 index counting the aforementioned fermion zero modes.

Finally, the non-trivial cobordism group in~\eqref{eq:SS32-cobordism-group} signals the existence of a \mbox{$(-1)$-form} global symmetry. In alignment with the No Global Symmetries Conjecture~\cite{Banks:1988yz,Banks:2010zn} and the Cobordism Conjecture~\cite{McNamara:2019rup} of the Swampland Program~\cite{Vafa:2005ui}, such symmetries should be either gauged or broken by objects in the theory, along the lines of~\cite{McNamara:2019rup,Blumenhagen:2021nmi,Blumenhagen:2022bvh}. One could then speculate that said \mbox{$(-1)$-form} global symmetry is gauged by the HO~instanton, but further work is required to firmly establish this point.

\section{Heterotic disks and D-instantons}
\label{sec:hetdisks}

We now argue that the instanton described above gives rise to $\mathcal{O}\left( e^{-1/g_s} \right)$ effects in the HO~theory. First, let us review Shenker's general prediction for such effects and how it is substantiated in the Type~I and~II theories.

In \emph{closed} string perturbation theory, a general amplitude has a genus expansion
\begin{equation}
    \mc{A} \simeq \sum_{n=0}^\infty a_n g_s^{2n-2} \fstop
\label{eq:stringpert}
\end{equation}
At large $n$, the coefficients $\{a_n\}_{n \in \mathbb{Z}_{\geq 0}}$ diverge factorially due to the integral over the genus $n$ Riemann surface moduli space~\cite{Gross:1988ib}. In fact, Shenker~\cite{Shenker:1990uf} determined the leading order behavior  $a_n\simeq C^{-2n}(2n!)$ at large $n$, where $C$ is a constant.

This divergence implies that~\eqref{eq:stringpert} is an asymptotic series with vanishing radius of convergence. To define the full non-perturbative amplitude, \eqref{eq:stringpert} must be supplemented by additional contributions arising from non-trivial saddle points in the path integral. The strength of the \textit{first} saddle is encoded in the divergence of the coefficients $\{a_n\}_{n \in \mathbb{Z}_{\geq 0}}$. Attempting a Borel resummation of \eqref{eq:stringpert} with Shenker's estimate reveals a leading order non-perturbative ambiguity of $\mathcal{O} \left( e^{-C/g_s} \right)$.

The physical origin of such saddles in Type~I and Type~II string theory is understood\,---\,\mbox{D$p$-branes} have tensions $\tau_{p} \propto g_s^{-1}$, meaning that Euclidean branes wrapping $(p+1)$-cycles of a compactification manifold contribute the required terms. In 10D, only \mbox{D$(-1)$-branes} are relevant. To make the connection more concrete, we turn to the path integral discussion of~\cite{Polchinski:1994fq}.\footnote{See also~\cite{Green:1994iw,Green:1995mu,Green:1995my,Green:1997tv}.}

In a D-dimensional theory, the contribution of \mbox{D$(-1)$-branes} to an amplitude stems from a path integral piece with the general form 
\begin{equation}
    \sum_{N=0}^\infty\prod_{i=1}^N\int \left[ d^{D}X_i \right] \sum_{n_i=0}^\infty\big(\cdots\big)\fstop
\label{eq:instantonpath}
\end{equation}
Here $X_i$ denotes the spacetime position of the $i$-th instanton, and the ellipsis corresponds to the rest of the path integral term, including vertex operators. The various parts of~\eqref{eq:instantonpath} prescribe that, for a fixed number of instantons, one should sum over an arbitrary number of worldsheets ending on them, integrate over their spacetime positions and, finally, sum over an arbitrary number of instantons. Restricting ourselves to the single-instanton terms,~\eqref{eq:instantonpath} includes a sum over $m$ disk diagrams that are disconnected from the rest of the amplitude. Since the Euler number of the disk is $\chi=1$, each disk is weighted by $g_s^{-1}$. Accounting for an $m!$ symmetry factor, these disconnected disks sum to an exponential, yielding precisely a contribution to the amplitude proportional to $e^{-C/g_s}$.

Synthesizing the above, Shenker's prediction amounts to the statement that the Type~I and Type~II closed string perturbation theories must be supplemented by an open string sector, which includes strings with endpoints on a \mbox{D$(-1)$-brane} responsible for the $\mathcal{O} \left( e^{-1/g_{s}} \right)$ effects in 10D. This conclusion relies on arguments agnostic to the amount of supersymmetry, and therefore holds also for the 26D bosonic string.

We propose that this logic extends to the HO~theory and its instanton\,---\,the HO~theory must contain a \mbox{$(-1)$-brane} with open HO~strings ending on it. However, at first glance, an open string sector appears antithetic to the very notion of heterotic CFTs. The variation of the worldsheet action for an \emph{open} heterotic string is
\begin{equation}
    \delta S_{\Sigma} = \frac{1}{2\pi} \left. \int d\tau\, \left( \lambda^a\delta\lambda_a - \psi^\mu \delta \psi_\mu \right) \right|_{\sigma=0}^{\sigma=\ell}\,.
\label{eq:hetendpoints}
\end{equation}
Here $\{ \lambda^{a} \}_{a \in \{1, \dotsc ,32\}}$ constitute the left-moving current algebra and $\{ \psi^{\mu} \}_{\mu \in \{ 0, \dotsc, 9 \}}$ are the right-moving superpartners of the worldsheet bosons. The cancellation of \eqref{eq:hetendpoints} requires that $\lambda^i = \pm \psi^i$ for all $i$ at the endpoints of the string. However, this condition \emph{cannot} be satisfied in heterotic theories due to the asymmetric CFT field content.

Despite this difficulty, open heterotic strings in the HO~theory were shown to exist in Lorentzian spacetime~\cite{Polchinski:2005bg}. The key to their consistency lies in the \mbox{0-branes} present at the endpoints of the HO string. Critically, the $S^{8}$ enclosing a \mbox{0-brane} supports a vector bundle with $\mathrm{SemiSpin}(32)$-structure in the adjoint representation and associated with a non-trivial homotopy class of $\pi_7(\HO) \cong \ints$.  We can choose this vector bundle such that its structure group reduces to a subgroup of $\HO$ with $\mathfrak{so}(8)$ Lie algebra. In this background, the gauginos give rise to spacetime zero modes $\{ \Lambda^{b} \}_{b \in \{ 1, \dotsc, 24 \}}$ that transform under the $\mathbf{24}$ dimensional representation of {the gauge subgroup with trivial bundle.\footnote{\label{foot:representations}The $\mathbf{492}$ faithful representation of $\mathrm{PSO}(32)$ lifts to $\mathrm{SemiSpin}(32)$. The embedding $\mathfrak{so}(32) \supset \mathfrak{so}(8) \oplus \mathfrak{so}(24)$ leads to the branching rule $\mathbf{496} = (\mathbf{28},\mathbf{1}) \oplus (\mathbf{1},\mathbf{276}) \oplus (\mathbf{8_{v}},\mathbf{24})$ \cite{Yamatsu:2015npn}, the r.h.s.\ corresponding to a representation of $(\mathrm{Spin}(8) \times \mathrm{Spin}(24))/(\mathbb{Z}_{2} \times \mathbb{Z}_{2})$, see~\cite{McInnes:1999va,McInnes:1999pt} for details on the global structure of the subgroups of $\mathrm{SemiSpin}(32)$.} These zero modes can also be enumerated via the Atiyah-Singer index theorem applied to the twisted Dirac operator defined on the enclosing sphere $S^8$.  The proposal of~\cite{Polchinski:2005bg} is that these zero modes latch onto the endpoints of the HO~string and satisfy $\Lambda^b = \pm\lambda^b$ for $b \in \{ 1, \dotsc, 24 \}$, with the remaining 8 current algebra fermions matched to the 8 (physical gauge) fermions $\{ \psi^\mu \}_{\mu \in \{2, \dotsc, 9\}}$.

We now extend this logic to Euclidean spacetime and propose that a similar mechanism exists to ensure the consistency of HO~endpoints on the \mbox{$(-1)$-brane}. As described above, the HO~instanton is characterized by a gauge profile associated with the non-trivial class of $\pi_9(\HO)\cong \ints_2$.  As in the case of the Type~I instanton discussed in~\cite{Witten:1998cd}, and in analogy with the preceding discussion, we can choose the vector bundle such that its structure group reduces to a subgroup with $\mathfrak{so}(10)$ Lie algebra. The mod 2 Atiyah-Singer index \mbox{theorem \cite{PMIHES_1969,Atiyah:1970ws,Atiyah:1971rm}} ensures that the number of gaugino zero modes in such a background is $1 \mod 2$. Furthermore, each of these zero modes $\{\Lambda^{b'}\}_{b' \in \{ 1, \dotsc, 22 \}}$ transform as a $\mathbf{22}$ dimensional representation of the gauge subgroup with trivial bundle.\footnote{The same considerations as in \cref{foot:representations} hold \textit{mutatis mutandis} for this case.} We then patch up the inconsistency of the heterotic CFT by demanding that $\Lambda^{b'} = \pm \lambda^{b'}$, for $b' \in \{ 1, \dotsc, 22 \}$, at the string endpoints. The remaining 10 current algebra fermions are handled by all $\{\psi^{\mu}\}_{\mu \in \{0, \dotsc, 9\}}$, since here we do not employ physical gauge.

The above suggests indeed that open HO~strings can end on the instanton due to a form of inflow for spacetime degrees of freedom onto the worldsheet.  Furthermore, the path integral argument ensures that disconnected disks appear. The key question remaining is: What should the action of this instanton be? If we consider an HO~disk diagram with endpoints on the instanton, the Euler number of the disk suggests that the action is proportional to $g_s^{-1}$. Such a scaling, paired with the path integral argument, would provide a precise realization of Shenker's prediction in the HO~theory. This inverse $g_s$ scaling is of course the universal feature of \mbox{D-branes} in the Type~I and Type~II theories. It is tempting to conclude the same holds in heterotic theories, but our situation is far more subtle\,---\,in matching the spacetime fermion zero modes with the worldsheet fields, we have gone beyond the usual paradigm of worldsheet (B)CFTs. 

Nonetheless, we argue that this naive answer appears correct. First, it seems that the $g_{s}^{-1}$ scaling is universal in the HO~theory as well. The heterotic ``D-string" from \cref{sec:inheritance} has a tension proportional to~$g_s^{-1}$, which is necessary to match the S-duality arguments of~\cite{Silverstein:1996xp}. Furthermore, the tentative (gauge) \mbox{0-brane} in the 9D HE~theory, discussed below, also follows this scaling relation. This suggests that HO~disk diagrams, just like their Type~I and Type~II counterparts, should also be associated with a $g_s^{-1}$ scaling. Secondly, we can motivate the scaling by appealing to the known results of~\cite{Green:2016tfs}\,---\,if the HO~instanton does indeed give rise to the Eisenstein series in \eqref{eq:eisenstein}, then its action must necessarily scale as $g_s^{-1}$.

While these arguments are quite suggestive, they are not a proof. To settle the $g_s$ scaling, one must develop the tools necessary to \emph{calculate} the effect of heterotic disks. We leave this as a task for the future. For the present, we state that, provided one accepts the above arguments, heterotic disks account for Shenker effects in the HO~theory.

Finally, it is instructive to contrast the role of spacetime zero modes and the \mbox{$(-1)$-brane} in the Type~I and HO frames. In Type~I, the F-string can end on the \mbox{$(-1)$-brane} without issue due to the symmetric worldsheet CFT field content, but the spacetime fermion zero modes were argued to be necessary in order to remove the disconnected piece of the perturbative $\mathrm{O}(32)$ gauge group~\cite{Witten:1998cd}. Instead, the HO~frame knows the correct gauge group at the perturbative level, but the fermion zero modes are required for consistency of the F-string endpoints on the \mbox{$(-1)$-brane}.

\section{Other heterotic theories}
\label{sec:other-heterotic-theories}

We have explained the origin of Shenker effects in the HO~theory, but one 10D superstring theory remains: the HE~theory. Naively, the above discussion does not apply because $\pi_9(\mathrm{E}_8) = 0$ and, hence, there is no gauge configuration giving the spacetime fermion zero modes required for the consistency of HE~disk diagrams. This appears consistent with the absence of a tower of instanton corrections to the HE $R^{4}$-term in 10D \cite{Green:2016tfs}: The non-trivial gauge profile was crucial for the instantons we just identified as contributing to this term in the HO~theory, but these must disappear in the \mbox{T-dual} decompactification frame. A similar statement was made in~\cite{Polchinski:2005bg} forbidding the existence of open HE cosmic strings in Minkowski spacetime due to $\pi_7(\mathrm{E}_8) = 0$.

Nonetheless, Shenker's argument applies to the HO and HE~theories equally well; the lack of the effects it predicts for the latter appears problematic. A resolution can be found by drawing an analogy with the Type~IIA theory, which has a non-BPS, uncharged, and unstable \mbox{D$(-1)$-brane}. This object does not contribute to the Type~IIA $R^{4}$-term in 10D, but should contribute as an unstable saddle to some set of processes~\cite{Sen:2020cef}.

This motivates us to consider \emph{gravitational} configurations, that are potentially unstable, to justify HE~string endpoints. The inflow then occurs due to the combined zero modes of the 10D gauginos, dilatino, and gravitino in the non-trivial spacetime.

In Lorentzian spacetime, we can consider an open HE~cosmic string with endpoints on a \mbox{0-brane} associated with a non-trivial gravitational charge. The \mbox{0-brane} exists in a spacetime that supports fermionic zero modes for the gauginos, dilatino, and gravitino. Applying the index theorems for the differential operators acting on Weyl spinors and a Rarita-Schwinger field~\cite{Alvarez-Gaume:1984zlq} supported on an \mbox{8-dimensional} manifold surrounding one of the HE string endpoints, consistency requires
\begin{equation}
\begin{split}
    \pm 24n = \frac12\int_{M^8} \biggl[ &495\left[\hat{A}(M^8)\right]_{8}\\
    &\quad + \left[\hat{A}(M^8)\left( \tr e^{iR/2\pi} - 1 \right)\right]_{8} \biggr]\,,
\label{eq:gravmodes}
\end{split}
\end{equation}
where $n\in\ints$ and $R$ is the Riemann tensor suitably contracted with $\mathrm{SO}(8)$ generators in its fundamental representation.

We expect that a similar situation holds for gravi\-tational instantons in the HE~theory arising from \mbox{$(-1)$-branes}. Endpoints of the HE string on this object are made consistent by a Euclidean 10D analogue of~\eqref{eq:gravmodes}, but with the zero modes adding up to a multiple of 22. A potential source for such effects are the 10D Hitchin spheres~\cite{Hitchin:1974rbi}, the higher-dimensional cousins of Milnor's exotic spheres~\cite{Milnor:1969}. Heterotic supergravity has an even number of fermionic zero modes on such manifolds~\cite{Witten:1985xe}, making them candidates that could realize purely gravitational Shenker effects in the 10D HE~theory.

This completes our discussion of Shenker effects for the 10D superstring theories, but does not exhaust the landscape of heterotic theories. First, upon compactification the situation appears significantly enriched thanks to the possibility of breaking the heterotic gauge group or having Euclidean objects wrap the cycles of the compactification variety. Indeed, the results of~\cite{Green:2016tfs} indicate that, in the HE~theory on $S^{1}$ with an appropriate Wilson line, a \mbox{0-brane} (whose existence can now be supported by a non-trivial gauge profile) contributes Shenker effects to the 9D $R^{4}$-term via Euclidean worldlines wrapping the circle. We leave the study of open heterotic strings in compactifications to future work.

Secondly, Shenker's argument should apply to other heterotic theories beyond the HO and HE~theories. For example, in 10D we have the non-supersymmetric \mbox{$\mathrm{SO}(16) \times \mathrm{SO}(16)$} heterotic string~\cite{Alvarez-Gaume:1986ghj,Dixon:1986iz}, whose global gauge group is $[(\mathrm{Spin(16) \times \mathrm{Spin}(16)})/\mathbb{Z}_{2}] \rtimes \mathbb{Z}_{2}$ \cite{McInnes:1999va,Fraiman:2023cpa}. Heterotic instantons similar to the ones discussed for the HO~theory should be possible. Studying the lower-dimensional analogues of this theory \cite{Baykara:2024tjr} would also be of interest. Another sector worth discussing is that populated by the non-critical string theories. The analogue of \mbox{Polchinski's} long string was considered in \cite{Seiberg:2005nk} for the 2D non-critical heterotic strings, finding that it plays a role in the cancellation of the gauge and gravitational anomalies of the twisted orbifold version of the HO~theory. It would be appealing to also discuss open heterotic strings and Shenker effects in the context of the supercritical $\mathrm{HO}^{+}$ and $\mathrm{HO}^{+/}$~theories~\cite{Hellerman:2004zm}. These are closely related to the conventional HO~theory, enabling, e.g., a K-theoretic description of its \mbox{NS5-brane} via tachyon condensation~\cite{Garcia-Etxebarria:2014txa}.

\section{Conclusions}
\label{sec:conclusions}

Heterotic non-perturbative effects of order $\mathcal{O} \left( e^{-1/g_{s}} \right)$ were predicted by Shenker \cite{Shenker:1990uf} and confirmed to exist by Green and Rudra \cite{Green:2016tfs}, but an explanation of the objects that give rise to them remained elusive. Here we have motivated that their origin is found in heterotic ``\mbox{D-instantons}," i.e.\ heterotic disk diagrams. These are highly non-perturbative configurations, forcing us to go well beyond the usual worldsheet (B)CFT paradigm by mixing worldsheet and spacetime degrees of freedom, expanding on the ideas of \cite{Polchinski:2005bg}.

The above discussion reveals several key lessons. The first is that those heterotic branes on which the F-string can end must feature an inflow mechanism for consistency, as exemplified by the \mbox{D-string} in~\cite{Hull:1997kt,Hull:1998he}, the \mbox{0-brane} in~\cite{Polchinski:2005bg} and the \mbox{$(-1)$-brane} discussed in this work.

Second, while heterotic disks lie outside the purview of typical worldsheet (B)CFTs, it appears one should nonetheless associate them with a scaling proportional to $g_s^{-1}$. This is required to match the contributions calculated in \cite{Green:2016tfs} arising from  the HO~\mbox{$(-1)$-brane} and 9D HE~(gauge) \mbox{0-brane}. The same scaling for the heterotic disks is required for the HO~\mbox{D-string} to match the duality arguments of \cite{Silverstein:1996xp}. This scaling agrees nicely with our intuition from the Euler number of a disk.

Finally, the reciprocal picture of an old string theory adage arises: It is common lore that a theory of open strings must contain closed strings due to the possibility of endpoint reconnection. Our discussion above indicates that Shenker's argument, drawn to its natural conclusion, implies that a theory of closed strings must also incorporate open strings.

The path to this conclusion involved a non-trivial confluence of several distinct topics, including dualities between string theories and quotients thereof, K-theory, fermion zero mode inflow arising from index theorems, and the theory of resurgence applied to string amplitudes.

It would be desirable to establish a consistent framework to calculate the effects of open heterotic strings from first principles, a question that merits future investigation. A more fundamental treatment of such effects may rely on non-perturbative approaches to quantum gravity, among which dualities, string field theory and matrix models have proven to be very useful elsewhere. Indeed, there exist hints of a connection between matrix models and heterotic Shenker effects \cite{Sethi:2013hra}.

Compactifications to lower dimensions and broken gauge groups may alter the way in which the heterotic Shenker effects are concretely realized. In 4D compactifications they may play a role in moduli stabilization \cite{Casas:1996zi,Binetruy:1996nx,Binetruy:1997vr,Barreiro:1997rp,Kaufman:2013pya,Gaillard:2007jr}, cosmology \cite{Leedom:2022zdm} and string phenomenology more broadly, making their study a crucial target.

\begin{acknowledgments}
    We thank N.\,Cribiori for collaboration on an early version of this work. We also thank V.\,Schomerus and A.\,Westphal for many helpful discussions. Finally, we thank R.\,Blumenhagen, N.\,Cribiori, C.\,Hull and T.\,Weigand for useful discussions and comments on the draft. \mbox{R.\,A.-G.} and J.\,M.\,L. are supported in part by Deutsche Forschungsgemeinschaft under Germany’s Excellence Strategy EXC 2121 Quantum Universe 390833306, by Deutsche Forschungsgemeinschaft through a German-Israeli Project Cooperation (DIP) grant ``Holography and the Swampland" and by Deutsche Forschungsgemeinschaft through the Collaborative Research Center SFB1624 ``Higher Structures, Moduli Spaces, and Integrability." N.\,R. is supported by a Leverhulme Trust Research Project Grant RPG-2021-423. J.\,M.\,L. thanks the Galileo Galilei Institute for Theoretical Physics for the hospitality and the INFN for partial support during the completion of part of this work. N.\,R. thanks DESY for hospitality during the initial stages of this work.
\end{acknowledgments}

\bibliography{references}

@PREAMBLE{
 "\providecommand{\noopsort}[1]{}" 
 # "\providecommand{\singleletter}[1]{#1}%" 
}

@article{Gross:1985fr,
    author = "Gross, David J. and Harvey, Jeffrey A. and Martinec, Emil J. and Rohm, Ryan",
    title = "{Heterotic String Theory. 1. The Free Heterotic String}",
    reportNumber = "PRINT-85-0203 (PRINCETON)",
    doi = "10.1016/0550-3213(85)90394-3",
    journal = "Nucl. Phys. B",
    volume = "256",
    pages = "253",
    year = "1985"
}

@article{Gross:1985rr,
    author = "Gross, David J. and Harvey, Jeffrey A. and Martinec, Emil J. and Rohm, Ryan",
    title = "{Heterotic String Theory. 2. The Interacting Heterotic String}",
    reportNumber = "Print-85-0694 (PRINCETON)",
    doi = "10.1016/0550-3213(86)90146-X",
    journal = "Nucl. Phys. B",
    volume = "267",
    pages = "75--124",
    year = "1986"
}

@article{Bergshoeff:2006bs,
    author = "Bergshoeff, Eric A. and Gibbons, Gary W. and Townsend, Paul K.",
    title = "{Open M5-branes}",
    eprint = "hep-th/0607193",
    archivePrefix = "arXiv",
    reportNumber = "DAMTP-2006-42, UG-06-06",
    doi = "10.1103/PhysRevLett.97.231601",
    journal = "Phys. Rev. Lett.",
    volume = "97",
    pages = "231601",
    year = "2006"
}

@article{Kaidi:2023tqo,
    author = "Kaidi, Justin and Ohmori, Kantaro and Tachikawa, Yuji and Yonekura, Kazuya",
    title = "{Nonsupersymmetric Heterotic Branes}",
    eprint = "2303.17623",
    archivePrefix = "arXiv",
    primaryClass = "hep-th",
    doi = "10.1103/PhysRevLett.131.121601",
    journal = "Phys. Rev. Lett.",
    volume = "131",
    number = "12",
    pages = "121601",
    year = "2023"
}

@article{Silverstein:1996xp,
    author = "Silverstein, Eva",
    title = "{Duality, compactification, and $e^{-1/\lambda}$ effects in the heterotic string theory}",
    eprint = "hep-th/9611195",
    archivePrefix = "arXiv",
    reportNumber = "RU-96-104",
    doi = "10.1016/S0370-2693(97)00098-1",
    journal = "Phys. Lett. B",
    volume = "396",
    pages = "91--96",
    year = "1997"
}

@article{vandeHeisteeg:2023dlw,
    author = "van de Heisteeg, Damian and Vafa, Cumrun and Wiesner, Max and Wu, David H.",
    title = "{Species scale in diverse dimensions}",
    eprint = "2310.07213",
    archivePrefix = "arXiv",
    primaryClass = "hep-th",
    doi = "10.1007/JHEP05(2024)112",
    journal = "JHEP",
    volume = "05",
    pages = "112",
    year = "2024"
}

@unpublished{McNamara:2019rup,
    author = "McNamara, Jacob and Vafa, Cumrun",
    title = "{Cobordism Classes and the Swampland}",
    eprint = "1909.10355",
    archivePrefix = "arXiv",
    primaryClass = "hep-th",
    month = "9",
    year = "2019"
}

@article{Blumenhagen:2021nmi,
    author = "Blumenhagen, Ralph and Cribiori, Niccol\`o",
    title = "{Open-closed correspondence of K-theory and cobordism}",
    eprint = "2112.07678",
    archivePrefix = "arXiv",
    primaryClass = "hep-th",
    reportNumber = "MPP-2021-201",
    doi = "10.1007/JHEP08(2022)037",
    journal = "JHEP",
    volume = "08",
    pages = "037",
    year = "2022"
}

@article{Blumenhagen:2022bvh,
    author = "Blumenhagen, Ralph and Cribiori, Niccol\`o and Kneissl, Christian and Makridou, Andriana",
    title = "{Dimensional Reduction of Cobordism and K-theory}",
    eprint = "2208.01656",
    archivePrefix = "arXiv",
    primaryClass = "hep-th",
    reportNumber = "MPP-2022-95",
    doi = "10.1007/JHEP03(2023)181",
    journal = "JHEP",
    volume = "03",
    pages = "181",
    year = "2023"
}

@article{Banks:1988yz,
    author = "Banks, Tom and Dixon, Lance J.",
    title = "{Constraints on String Vacua with Spacetime Supersymmetry}",
    reportNumber = "PUPT-1086, SCIPP-8805",
    doi = "10.1016/0550-3213(88)90523-8",
    journal = "Nucl. Phys. B",
    volume = "307",
    pages = "93--108",
    year = "1988"
}

@article{Banks:2010zn,
    author = "Banks, Tom and Seiberg, Nathan",
    title = "{Symmetries and Strings in Field Theory and Gravity}",
    eprint = "1011.5120",
    archivePrefix = "arXiv",
    primaryClass = "hep-th",
    doi = "10.1103/PhysRevD.83.084019",
    journal = "Phys. Rev. D",
    volume = "83",
    pages = "084019",
    year = "2011"
}

@article{Witten:1998cd,
    author = "Witten, Edward",
    title = "{D-branes and K-theory}",
    eprint = "hep-th/9810188",
    archivePrefix = "arXiv",
    reportNumber = "IASSNS-HEP-98-82",
    doi = "10.1088/1126-6708/1998/12/019",
    journal = "JHEP",
    volume = "12",
    pages = "019",
    year = "1998"
}

@article{Hull:1997kt,
    author = "Hull, C. M.",
    title = "{Gravitational duality, branes and charges}",
    eprint = "hep-th/9705162",
    archivePrefix = "arXiv",
    reportNumber = "QMW-PH-97-16, NI-97028-NQF",
    doi = "10.1016/S0550-3213(97)00501-4",
    journal = "Nucl. Phys. B",
    volume = "509",
    pages = "216--251",
    year = "1998"
}

@article{Hull:1998he,
    author = "Hull, C. M.",
    title = "{The Nonperturbative SO(32) heterotic string}",
    eprint = "hep-th/9812210",
    archivePrefix = "arXiv",
    reportNumber = "QMW-98-43",
    doi = "10.1016/S0370-2693(99)00802-3",
    journal = "Phys. Lett. B",
    volume = "462",
    pages = "271--276",
    year = "1999"
}

@article{Pantev:2016nze,
    author = "Pantev, T. and Sharpe, E.",
    title = "{Duality group actions on fermions}",
    eprint = "1609.00011",
    archivePrefix = "arXiv",
    primaryClass = "hep-th",
    doi = "10.1007/JHEP11(2016)171",
    journal = "JHEP",
    volume = "11",
    pages = "171",
    year = "2016"
}

@article{Tachikawa:2018njr,
    author = "Tachikawa, Yuji and Yonekura, Kazuya",
    title = "{Why are fractional charges of orientifolds compatible with Dirac quantization?}",
    eprint = "1805.02772",
    archivePrefix = "arXiv",
    primaryClass = "hep-th",
    reportNumber = "IPMU-18-0067",
    doi = "10.21468/SciPostPhys.7.5.058",
    journal = "SciPost Phys.",
    volume = "7",
    number = "5",
    pages = "058",
    year = "2019"
}

@inproceedings{Dabholkar:1997zd,
    author = "Dabholkar, Atish",
    title = "{Lectures on orientifolds and duality}",
    booktitle = "{ICTP Summer School in High-Energy Physics and Cosmology}",
    eprint = "hep-th/9804208",
    archivePrefix = "arXiv",
    reportNumber = "TIFR-TH-98-13",
    pages = "128--191",
    month = "6",
    year = "1997"
}

@article{Bergshoeff:1998re,
    author = "Bergshoeff, E. and Eyras, E. and Halbersma, R. and van der Schaar, J. P. and Hull, C. M. and Lozano, Y.",
    editor = "Lust, D. and Otto, H. J.",
    title = "{Space-time filling branes and strings with sixteen supercharges}",
    eprint = "hep-th/9812224",
    archivePrefix = "arXiv",
    reportNumber = "UG-15-98, QMW-PH-98-39, SPIN-1998-14",
    doi = "10.1016/S0550-3213(99)00483-6",
    journal = "Nucl. Phys. B",
    volume = "564",
    pages = "29--59",
    year = "2000"
}

@article{Adams:2010zy,
    author = "Adams, Allan and DeWolfe, Oliver and Taylor, Washington",
    title = "{String universality in ten dimensions}",
    eprint = "1006.1352",
    archivePrefix = "arXiv",
    primaryClass = "hep-th",
    reportNumber = "COLO-HEP-554, MIT-CTP-4155",
    doi = "10.1103/PhysRevLett.105.071601",
    journal = "Phys. Rev. Lett.",
    volume = "105",
    pages = "071601",
    year = "2010"
}

@article{Sen:1996na,
    author = "Sen, Ashoke",
    title = "{Duality and orbifolds}",
    eprint = "hep-th/9604070",
    archivePrefix = "arXiv",
    reportNumber = "MRI-PHY-96-12",
    doi = "10.1016/0550-3213(96)00291-X",
    journal = "Nucl. Phys. B",
    volume = "474",
    pages = "361--378",
    year = "1996"
}

@article{Sen:1996yy,
    author = "Sen, Ashoke",
    editor = "Froehlich, J. and Rittenberg, V. and Schwimmer, A.",
    title = "{Unification of string dualities}",
    eprint = "hep-th/9609176",
    archivePrefix = "arXiv",
    reportNumber = "MRI-PHY-96-28",
    doi = "10.1016/S0920-5632(97)00409-X",
    journal = "Nucl. Phys. B Proc. Suppl.",
    volume = "58",
    pages = "5--19",
    year = "1997"
}

@article{Sen:1998ki,
    author = "Sen, Ashoke",
    title = "{Type I D-particle and its interactions}",
    eprint = "hep-th/9809111",
    archivePrefix = "arXiv",
    reportNumber = "MRI-PHY-P980960",
    doi = "10.1088/1126-6708/1998/10/021",
    journal = "JHEP",
    volume = "10",
    pages = "021",
    year = "1998"
}

@article{Frau:1999qs,
    author = "Frau, M. and Gallot, L. and Lerda, A. and Strigazzi, P.",
    title = "{Stable non-BPS D-branes in Type I string theory}",
    eprint = "hep-th/9903123",
    archivePrefix = "arXiv",
    reportNumber = "DFTT-14-99",
    doi = "10.1016/S0550-3213(99)00624-0",
    journal = "Nucl. Phys. B",
    volume = "564",
    pages = "60--85",
    year = "2000"
}

@article{Frau:2000hq,
    author = "Frau, M. and Gallot, L. and Lerda, A. and Strigazzi, P.",
    editor = "Bergshoeff, E. A. and Ceresole, Anna and Kounnas, C. and Lust, D. and Sevrin, A.",
    title = "{D-branes in Type I String Theory}",
    eprint = "hep-th/0012167",
    archivePrefix = "arXiv",
    reportNumber = "DFTT-49-2000",
    doi = "10.1002/1521-3978(200105)49:4/6<503::AID-PROP503>3.3.CO;2-T",
    journal = "Fortsch. Phys.",
    volume = "49",
    pages = "503--510",
    year = "2001"
}

@article{Witten:1995ex,
    author = "Witten, Edward",
    title = "{String theory dynamics in various dimensions}",
    eprint = "hep-th/9503124",
    archivePrefix = "arXiv",
    reportNumber = "IASSNS-HEP-95-18",
    doi = "10.1201/9781482268737-32",
    journal = "Nucl. Phys. B",
    volume = "443",
    pages = "85--126",
    year = "1995"
}

@article{Polchinski:1995df,
    author = "Polchinski, Joseph and Witten, Edward",
    title = "{Evidence for Heterotic - Type I String Duality}",
    eprint = "hep-th/9510169",
    archivePrefix = "arXiv",
    reportNumber = "IASSNS-HEP-95-81, NSF-ITP-95-135",
    doi = "10.1016/0550-3213(95)00614-1",
    journal = "Nucl. Phys. B",
    volume = "460",
    pages = "525--540",
    year = "1996"
}

@article{Balthazar:2019rnh,
    author = "Balthazar, Bruno and Rodriguez, Victor A. and Yin, Xi",
    title = "{ZZ instantons and the non-perturbative dual of c = 1 string theory}",
    eprint = "1907.07688",
    archivePrefix = "arXiv",
    primaryClass = "hep-th",
    doi = "10.1007/JHEP05(2023)048",
    journal = "JHEP",
    volume = "05",
    pages = "048",
    year = "2023"
}

@article{Balthazar:2019ypi,
    author = "Balthazar, Bruno and Rodriguez, Victor A. and Yin, Xi",
    title = "{Multi-instanton calculus in c = 1 string theory}",
    eprint = "1912.07170",
    archivePrefix = "arXiv",
    primaryClass = "hep-th",
    doi = "10.1007/JHEP05(2023)050",
    journal = "JHEP",
    volume = "05",
    pages = "050",
    year = "2023"
}

@article{Sen:2020cef,
    author = "Sen, Ashoke",
    title = "{D-instanton Perturbation Theory}",
    eprint = "2002.04043",
    archivePrefix = "arXiv",
    primaryClass = "hep-th",
    doi = "10.1007/JHEP08(2020)075",
    journal = "JHEP",
    volume = "08",
    pages = "075",
    year = "2020"
}

@article{Sethi:2013hra,
    author = "Sethi, Savdeep",
    title = "{A New String in Ten Dimensions?}",
    eprint = "1304.1551",
    archivePrefix = "arXiv",
    primaryClass = "hep-th",
    doi = "10.1007/JHEP09(2013)149",
    journal = "JHEP",
    volume = "09",
    pages = "149",
    year = "2013"
}

@article{Casas:1996zi,
    author = "Casas, J. A.",
    title = "{The generalized dilaton supersymmetry breaking scenario}",
    eprint = "hep-th/9605180",
    archivePrefix = "arXiv",
    reportNumber = "SCIPP-96-20, IEM-FT-129-96",
    doi = "10.1016/0370-2693(96)00821-0",
    journal = "Phys. Lett. B",
    volume = "384",
    pages = "103--110",
    year = "1996"
}

@article{Binetruy:1996nx,
    author = "Binetruy, Pierre and Gaillard, Mary K. and Wu, Yi-Yen",
    title = "{Modular invariant formulation of multi-gaugino and matter condensation}",
    eprint = "hep-th/9611149",
    archivePrefix = "arXiv",
    reportNumber = "LBL-39608, LBNL-39608, UCB-PTH-96-54",
    doi = "10.1016/S0550-3213(97)00162-4",
    journal = "Nucl. Phys. B",
    volume = "493",
    pages = "27--55",
    year = "1997"
}

@article{Binetruy:1997vr,
    author = "Binetruy, Pierre and Gaillard, Mary K. and Wu, Yi-Yen",
    title = "{Supersymmetry breaking and weakly versus strongly coupled string theory}",
    eprint = "hep-th/9702105",
    archivePrefix = "arXiv",
    reportNumber = "LBL-39744, LBNL-39744, UCB-PTH-96-61",
    doi = "10.1016/S0370-2693(97)00989-1",
    journal = "Phys. Lett. B",
    volume = "412",
    pages = "288--295",
    year = "1997"
}

@article{Barreiro:1997rp,
    author = "Barreiro, T. and de Carlos, B. and Copeland, Edmund J.",
    title = "{On nonperturbative corrections to the K{\"a}hler potential}",
    eprint = "hep-ph/9712443",
    archivePrefix = "arXiv",
    reportNumber = "SUSX-TH-97-024, IEM-FT-169-97",
    doi = "10.1103/PhysRevD.57.7354",
    journal = "Phys. Rev. D",
    volume = "57",
    pages = "7354--7360",
    year = "1998"
}

@article{Kaufman:2013pya,
    author = "Kaufman, Bryan L. and Nelson, Brent D. and Gaillard, Mary K.",
    title = {{Mirage models confront the LHC: K\"ahler-stabilized heterotic string theory}},
    eprint = "1303.6575",
    archivePrefix = "arXiv",
    primaryClass = "hep-ph",
    reportNumber = "UCB-PTH-13-02",
    doi = "10.1103/PhysRevD.88.025003",
    journal = "Phys. Rev. D",
    volume = "88",
    number = "2",
    pages = "025003",
    year = "2013"
}

@article{Gaillard:2007jr,
    author = "Gaillard, Mary K. and Nelson, Brent D.",
    title = "{K{\"a}hler stabilized, modular invariant heterotic string models}",
    eprint = "hep-th/0703227",
    archivePrefix = "arXiv",
    reportNumber = "UCB-PTH-07-05, NSF-KITP-07-27",
    doi = "10.1142/S0217751X07036439",
    journal = "Int. J. Mod. Phys. A",
    volume = "22",
    pages = "1451--1588",
    year = "2007"
}

@article{Leedom:2022zdm,
    author = "Leedom, Jacob M. and Righi, Nicole and Westphal, Alexander",
    title = "{Heterotic de Sitter beyond modular symmetry}",
    eprint = "2212.03876",
    archivePrefix = "arXiv",
    primaryClass = "hep-th",
    reportNumber = "DESY 22-173, KCL-PH-TH/2022-54",
    doi = "10.1007/JHEP02(2023)209",
    journal = "JHEP",
    volume = "02",
    pages = "209",
    year = "2023"
}

@article{Alvarez-Gaume:1984zlq,
    author = "Alvarez-Gaume, Luis and Ginsparg, Paul H.",
    editor = "Salam, A. and Sezgin, E.",
    title = "{The Structure of Gauge and Gravitational Anomalies}",
    reportNumber = "HUTP-84/A016",
    doi = "10.1016/0003-4916(85)90087-9",
    journal = "Annals Phys.",
    volume = "161",
    pages = "423",
    year = "1985",
    note = "[Erratum: Annals Phys. 171, 233 (1986)]"
}

@unpublished{Fraiman:2023cpa,
    author = "Fraiman, Bernardo and Gra\~na, Mariana and Parra De Freitas, H\'ector and Sethi, Savdeep",
    title = "{Non-Supersymmetric Heterotic Strings on a Circle}",
    eprint = "2307.13745",
    archivePrefix = "arXiv",
    primaryClass = "hep-th",
    month = "7",
    year = "2023"
}

@unpublished{Baykara:2024tjr,
    author = "Baykara, Zihni Kaan and Tarazi, Houri-Christina and Vafa, Cumrun",
    title = "{New Non-Supersymmetric Tachyon-Free Strings}",
    eprint = "2406.00185",
    archivePrefix = "arXiv",
    primaryClass = "hep-th",
    month = "5",
    year = "2024"
}

@article{Seiberg:2005nk,
    author = "Seiberg, Nathan",
    title = "{Long strings, anomaly cancellation, phase transitions, T-duality and locality in the 2d heterotic string}",
    eprint = "hep-th/0511220",
    archivePrefix = "arXiv",
    doi = "10.1088/1126-6708/2006/01/057",
    journal = "JHEP",
    volume = "01",
    pages = "057",
    year = "2006"
}

@unpublished{Hellerman:2004zm,
    author = "Hellerman, Simeon",
    title = "{On the landscape of superstring theory in $\mathrm{D} > 10$}",
    eprint = "hep-th/0405041",
    archivePrefix = "arXiv",
    month = "5",
    year = "2004"
}

@article{Garcia-Etxebarria:2014txa,
    author = "Garc\'\i{}a-Etxebarria, I{\~}naki and Montero, Miguel and Uranga, Angel",
    title = "{Heterotic NS5-branes from closed string tachyon condensation}",
    eprint = "1405.0009",
    archivePrefix = "arXiv",
    primaryClass = "hep-th",
    reportNumber = "MPP-2014-171, IFT-UAM-CSIC-14-035, FTUAM-14-15",
    doi = "10.1103/PhysRevD.90.126002",
    journal = "Phys. Rev. D",
    volume = "90",
    number = "12",
    pages = "126002",
    year = "2014"
}

@article{McInnes:1999va,
    author = "McInnes, Brett",
    title = "{The Semispin groups in string theory}",
    eprint = "hep-th/9906059",
    archivePrefix = "arXiv",
    doi = "10.1063/1.532999",
    journal = "J. Math. Phys.",
    volume = "40",
    pages = "4699--4712",
    year = "1999"
}

@article{McInnes:1999pt,
    author = "McInnes, Brett",
    title = "{Gauge spinors and string duality}",
    eprint = "hep-th/9910100",
    archivePrefix = "arXiv",
    reportNumber = "NUS-MATH-761",
    doi = "10.1016/S0550-3213(00)00025-0",
    journal = "Nucl. Phys. B",
    volume = "577",
    pages = "439--460",
    year = "2000"
}

@unpublished{Yamatsu:2015npn,
    author = "Yamatsu, Naoki",
    title = "{Finite-Dimensional Lie Algebras and Their Representations for Unified Model Building}",
    eprint = "1511.08771",
    archivePrefix = "arXiv",
    primaryClass = "hep-ph",
    reportNumber = "OU-HET 886, KYUSHU-HET-216, OU-HET-886",
    month = "11",
    year = "2015"
}

@unpublished{SS32Cobordism,
    author = "Knei{\ss}l, Christian",
    title = "{Spin cobordism and the gauge group of Type I/heterotic string theory}",
    eprint = "2407.20333",
    archivePrefix = "arXiv",
    primaryClass = "hep-th",
    reportNumber = "MPP-2024-159",
    month = "7",
    year = "2024"
}

@inproceedings{Witten:2019bou,
    author = "Witten, Edward and Yonekura, Kazuya",
    title = "{Anomaly Inflow and the \mbox{$\eta$-Invariant}}",
    booktitle = "{The Shoucheng Zhang Memorial Workshop}",
    eprint = "1909.08775",
    archivePrefix = "arXiv",
    primaryClass = "hep-th",
    month = "9",
    year = "2019"
}

@article{Witten:1982fp,
    author = "Witten, Edward",
    editor = "Shifman, Mikhail A.",
    title = "{An SU(2) Anomaly}",
    doi = "10.1016/0370-2693(82)90728-6",
    journal = "Phys. Lett. B",
    volume = "117",
    pages = "324--328",
    year = "1982"
}

@article{Anderson:1967, 
    title={The structure of the spin cobordism ring}, 
    volume={86}, 
    DOI={10.2307/1970690}, 
    number={2}, 
    journal={The Annals of Mathematics}, 
    author={Anderson, D. W. and Brown, E. H. and Peterson, F. P.}, 
    year={1967}, 
    month={Sep}, 
    pages={271}
}

@article{PMIHES_1969,
     author = {Atiyah, Michael F. and Singer, Isadore M.},
     title = {Index theory for skew-adjoint {Fredholm} operators},
     journal = {Publications Math\'ematiques de l'IH\'ES},
     pages = {5--26},
     publisher = {Institut des Hautes \'Etudes Scientifiques},
     volume = {37},
     year = {1969},
     mrnumber = {285033},
     zbl = {0194.55503},
     url = {http://www.numdam.org/item/PMIHES_1969__37__5_0/}
}

@article{Atiyah:1970ws,
    author = "Atiyah, M. F. and Singer, I. M.",
    title = "{The Index of Elliptic Operators: 4}",
    doi = "10.2307/1970756",
    journal = "Annals Math.",
    volume = "93",
    pages = "119--138",
    year = "1971"
}

@article{Atiyah:1971rm,
    author = "Atiyah, M. F. and Singer, I. M.",
    title = "{The Index of Elliptic Operators: 5}",
    doi = "10.2307/1970757",
    journal = "Annals Math.",
    volume = "93",
    pages = "139--149",
    year = "1971"
}

@article{Gross:1988ib,
    author = "Gross, David J. and Periwal, Vipul",
    title = "{String Perturbation Theory Diverges}",
    reportNumber = "PUPT-1090",
    doi = "10.1103/PhysRevLett.60.2105",
    journal = "Phys. Rev. Lett.",
    volume = "60",
    pages = "2105",
    year = "1988"
}

@inproceedings{Shenker:1990uf,
    author = "Shenker, Stephen H.",
    title = "{The Strength of nonperturbative effects in string theory}",
    booktitle = "{Cargese Study Institute: Random Surfaces, Quantum Gravity and Strings}",
    reportNumber = "RU-90-47, RU-90-047",
    pages = "809--819",
    month = "8",
    year = "1990"
}

@article{Polchinski:1994fq,
    author = "Polchinski, Joseph",
    title = "{Combinatorics of boundaries in string theory}",
    eprint = "hep-th/9407031",
    archivePrefix = "arXiv",
    reportNumber = "NSF-ITP-94-73",
    doi = "10.1103/PhysRevD.50.R6041",
    journal = "Phys. Rev. D",
    volume = "50",
    pages = "R6041--R6045",
    year = "1994"
}

@article{Polchinski:2005bg,
    author = "Polchinski, Joseph",
    title = "{Open heterotic strings}",
    eprint = "hep-th/0510033",
    archivePrefix = "arXiv",
    doi = "10.1088/1126-6708/2006/09/082",
    journal = "JHEP",
    volume = "09",
    pages = "082",
    year = "2006"
}

@article{Green:1994iw,
    author = "Green, Michael B.",
    title = "{Point-Like States for Type 2b Superstrings}",
    eprint = "hep-th/9403040",
    archivePrefix = "arXiv",
    reportNumber = "DAMTP-94-19",
    doi = "10.1016/0370-2693(94)91087-1",
    journal = "Phys. Lett. B",
    volume = "329",
    pages = "435--443",
    year = "1994"
}

@inproceedings{Green:1995mu,
    author = "Green, Michael B.",
    title = "{Boundary effects in string theory}",
    booktitle = "{STRINGS 95: Future Perspectives in String Theory}",
    eprint = "hep-th/9510016",
    archivePrefix = "arXiv",
    reportNumber = "CERN-TH-95-252",
    pages = "220--229",
    month = "3",
    year = "1995"
}

@article{Green:1995my,
    author = "Green, Michael B.",
    title = "{A Gas of D-instantons}",
    eprint = "hep-th/9504108",
    archivePrefix = "arXiv",
    reportNumber = "CERN-TH-95-78, CERN-TH-95-078",
    doi = "10.1016/0370-2693(95)00584-8",
    journal = "Phys. Lett. B",
    volume = "354",
    pages = "271--278",
    year = "1995"
}

@article{Green:1997tv,
    author = "Green, Michael B. and Gutperle, Michael",
    title = "{Effects of D-instantons}",
    eprint = "hep-th/9701093",
    archivePrefix = "arXiv",
    reportNumber = "DAMTP-96-104",
    doi = "10.1016/S0550-3213(97)00269-1",
    journal = "Nucl. Phys. B",
    volume = "498",
    pages = "195--227",
    year = "1997"
}

@article{Green:2016tfs,
    author = "Green, Michael B. and Rudra, Arnab",
    title = "{Type I/heterotic duality and M-theory amplitudes}",
    eprint = "1604.00324",
    archivePrefix = "arXiv",
    primaryClass = "hep-th",
    reportNumber = "DAMTP-2016-22",
    doi = "10.1007/JHEP12(2016)060",
    journal = "JHEP",
    volume = "12",
    pages = "060",
    year = "2016"
}

@article{Milnor:1969,
 ISSN = {0003486X, 19398980},
 URL = {http://www.jstor.org/stable/1969983},
 author = {John Milnor},
 journal = {Annals of Mathematics},
 number = {2},
 pages = {399--405},
 publisher = {[Annals of Mathematics, Trustees of Princeton University on Behalf of the Annals of Mathematics, Mathematics Department, Princeton University]},
 title = {On Manifolds Homeomorphic to the 7-Sphere},
 urldate = {2024-07-19},
 volume = {64},
 year = {1956}
}

@article{Witten:1985xe,
    author = "Witten, Edward",
    editor = "Salam, A. and Sezgin, E.",
    title = "{Global gravitational anomalies}",
    reportNumber = "PRINT-85-0246 (PRINCETON)",
    doi = "10.1007/BF01212448",
    journal = "Commun. Math. Phys.",
    volume = "100",
    pages = "197",
    year = "1985"
}

@unpublished{Vafa:2005ui,
    author = "Vafa, Cumrun",
    title = "{The String landscape and the swampland}",
    eprint = "hep-th/0509212",
    archivePrefix = "arXiv",
    reportNumber = "HUTP-05-A043",
    month = "9",
    year = "2005"
}

@article{Hitchin:1974rbi,
    author = "Hitchin, Nigel",
    title = "{Harmonic Spinors}",
    doi = "10.1016/0001-8708(74)90021-8",
    journal = "Adv. Math.",
    volume = "14",
    number = "1",
    pages = "1--55",
    year = "1974"
}

@article{Alvarez-Gaume:1986ghj,
    author = "Alvarez-Gaume, Luis and Ginsparg, Paul H. and Moore, Gregory W. and Vafa, C.",
    title = "{An O(16) x O(16) Heterotic String}",
    reportNumber = "HUTP-86/A013",
    doi = "10.1016/0370-2693(86)91524-8",
    journal = "Phys. Lett. B",
    volume = "171",
    pages = "155--162",
    year = "1986"
}

@article{Dixon:1986iz,
    author = "Dixon, Lance J. and Harvey, Jeffrey A.",
    editor = "Schellekens, B.",
    title = "{String Theories in Ten-Dimensions Without Space-Time Supersymmetry}",
    reportNumber = "PRINT-86-0244 (PRINCETON)",
    doi = "10.1016/0550-3213(86)90619-X",
    journal = "Nucl. Phys. B",
    volume = "274",
    pages = "93--105",
    year = "1986"
}

@article{Derrick:1964ww,
    author = "Derrick, G. H.",
    title = "{Comments on Nonlinear Wave Equations as Models for Elementary Particles}",
    doi = "10.1063/1.1704233",
    journal = "J. Math. Phys.",
    volume = "5",
    pages = "1252--1254",
    year = "1964"
}

@article{Hull:1994ys,
    author = "Hull, C. M. and Townsend, P. K.",
    title = "{Unity of superstring dualities}",
    eprint = "hep-th/9410167",
    archivePrefix = "arXiv",
    reportNumber = "QMW-94-30, DAMTP-R-94-33",
    doi = "10.1201/9781482268737-24",
    journal = "Nucl. Phys. B",
    volume = "438",
    pages = "109--137",
    year = "1995"
}

\end{document}